\documentclass[twocolumn,showpacs,preprintnumbers,amsmath,amssymb]{revtex4}

\usepackage{graphicx} % Include figure files
\usepackage{dcolumn} % Align table columns on decimal point
\usepackage{bm} % bold math

\begin{document}

\title{Magnetophoresis of Flexible DNA-based Dumbbell Structures}

\author{B.\ Babi\'c}
\author{R.\ Ghai}
\author{K. Dimitrov}
\email{k.dimitrov@uq.edu.au}
\affiliation{Australian Institute for
Bioengineering and Nanotechnology, Building 75 - Cnr of College
and Cooper Road, The University of Queensland, Brisbane, QLD, 4072
Australia }
\date{\today}% It is always \today, today,

\begin{abstract}

Controlled movement and manipulation of magnetic micro and
nanostructures using magnetic forces can give rise to important
applications in biomedecine, diagnostics and immunology. We report
controlled magnetophoresis and stretching, in aqueous solution, of
a DNA-based dumbbell structure containing magnetic and diamagnetic
microspheres. The velocity and stretching of the dumbbell were
experimentally measured and correlated with a theoretical model
based on the forces acting on individual magnetic beads or the
entire dumbbell structures. The results show that precise and
predictable manipulation of dumbbell structures is achievable and
can potentially be applied to immunomagnetic cell separators.

\end{abstract}

% Biological fluid dynamics, 47.63.–b
% Magnetic devices, 85.70.–w
% DNA, 87.14.Gg
% Fluidics, 47.85.Np

\maketitle

% ------------------------------------------------------------------
% Main text
% -------------------------------------------------------------------

%%%%%%%%%
% Intro %
%%%%%%%%%
Magnetic micro- and nanoparticles find widespread applications in
biotechnology, e.g. separation of cells and biomolecules.
Commercially available, micrometer-sized magnetic beads (MBs) are
polymer spheres containing homogenously dispersed
superparamagnetic iron oxide nanoparticles~\cite{Fonnum}. To make
them suitable for biological uses their surface is typically
functionalized with high affinity binding proteins e.g.
Streptavidin or protein A. Such affinity functionalized beads can
then be bound specifically to cells or biomolecules for magnetic
separation from biological samples (magnetic pulldown). In
contrast to such bulk magnetic separation, controlled manipulation
and movement (magnetophoresis) of individual magnetic beads
requires more intricate control over external magnetic fields.
There have been efforts in the last \mbox{$20$} years to optimize
magnetophoretic conditions  for this purpose~\cite{Inglis1,Chiou},
however, applications involving magnetophoresis are still scarce
and almost exclusively tested on individual
MBs~\cite{Fonnum,Pamme,Watarai}.

%%%%%%%%%%%%%%%%%%%%%%%%
% Theory %
%%%%%%%%%%%%%%%%%%%%%%%%
General theoretical basis for MB magnetophoresis can be derived by
accounting for the forces acting on the MB. In the presence of an
inhomogeneous magnetic field, a magnetic force
\mbox{$\mathbf{F}_m=\nabla(\mathbf{m}\cdot\mathbf{B})$} is exerted
on a MB. Here, \mbox{$\mathbf{m}$} is the magnetization of a bead
in the magnetic field \mbox{$\mathbf{B}$}. In the conditions used
in this study the magnetization of a MB is unsaturated (the
saturation magnetic field is \mbox{$B_s\approx0.5$\,T}), and hence
the magnetic response of the bead is described  as a linear
function of the volumetric magnetic susceptibility,
\mbox{$\chi_b$}. Therefore, the magnetic  force becomes:
\begin{equation}\label{magneticforcesus}
\mathbf{F}_m=\frac{V(\chi_b-\chi_s)}{\mu_0}\mathbf{B}(\nabla \cdot
\mathbf{B}),
\end{equation}
where $V$ is MB's volume, \mbox{$\chi_s$} is the volume magnetic
susceptibility of the surrounding aqueous solution and
\mbox{$\mu_0$} is the vacuum magnetic permeability. In addition to
this magnetic force, MBs moving in a fluid also experience a
counteracting viscosity-related drag force. For a spherical
particle in laminar flow, the drag force is
\mbox{$\mathbf{F}_d=6r\pi\eta\mathbf{v}$}, where $\eta$ is
viscosity of a fluid, and \textbf{v} and $r$ stand for velocity
and radius of a bead, respectively. Thus, the net force acting on
a MB is given as:

\begin{equation}\label{totalforce}
\mathbf{F}=\mathbf{F}_m-\mathbf{F}_d=\frac{V(\chi_b-\chi_s)}{\mu_0}(\mathbf{B}\nabla
\cdot\mathbf{B})-6r\pi\eta \mathbf{v}.
\end{equation}

%%%%%%%%%%%%%%%%%%%%%%%%%%%%%%%%%%%%%%%%%%%
% Sample preparation%
%%%%%%%%%%%%%%%%%%%%%%%%%%%%%%%%%%%%%%%%%%%

Magnetic separations of specific immune cell types and populations
have found widespread use in research, diagnostic and therapeutic
uses. Immunochemical methods have been developed for attachment of
magnetic beads to the cell surfaces, typically followed by
magnetic separation~\cite{Thiel}. Frequent problem encountered in
this type of "positive" immunomagnetic sorting is that the immune
cell can "sense" the magnetic bead, thus triggering immunogenic
responses such as activation and internalization~\cite{Oren}. This
often necessitates the use of an alternative, "negative" sorting
technique that targets for removal all cell types with the
exception of the cell population of interest, which remains
unattached~\cite{Cotter}. Negative sorting is an expensive and
cumbersome proposition, as it requires the use of large quantities
and diverse panels of magnetic beads directed at many divergent
cell types.

We have reasoned that formation of dumbbell structures by
attaching magnetic beads to immune cells via polymeric linkers
would provide sufficient separation between the MB and the cell
surface to minimize activation and internalization, and thus would
be an attractive alternative to negative immunomagnetic sorting.
In this letter, we report on an experimental study of controlled
movement of model structures consisting of magnetic beads and
latex diamagnetic spheres, joined via DNA linkers, as a physical
model for magnethophoretic separation of immunological cell types
via dumbbells.

\begin{figure}
\includegraphics[width=75mm]{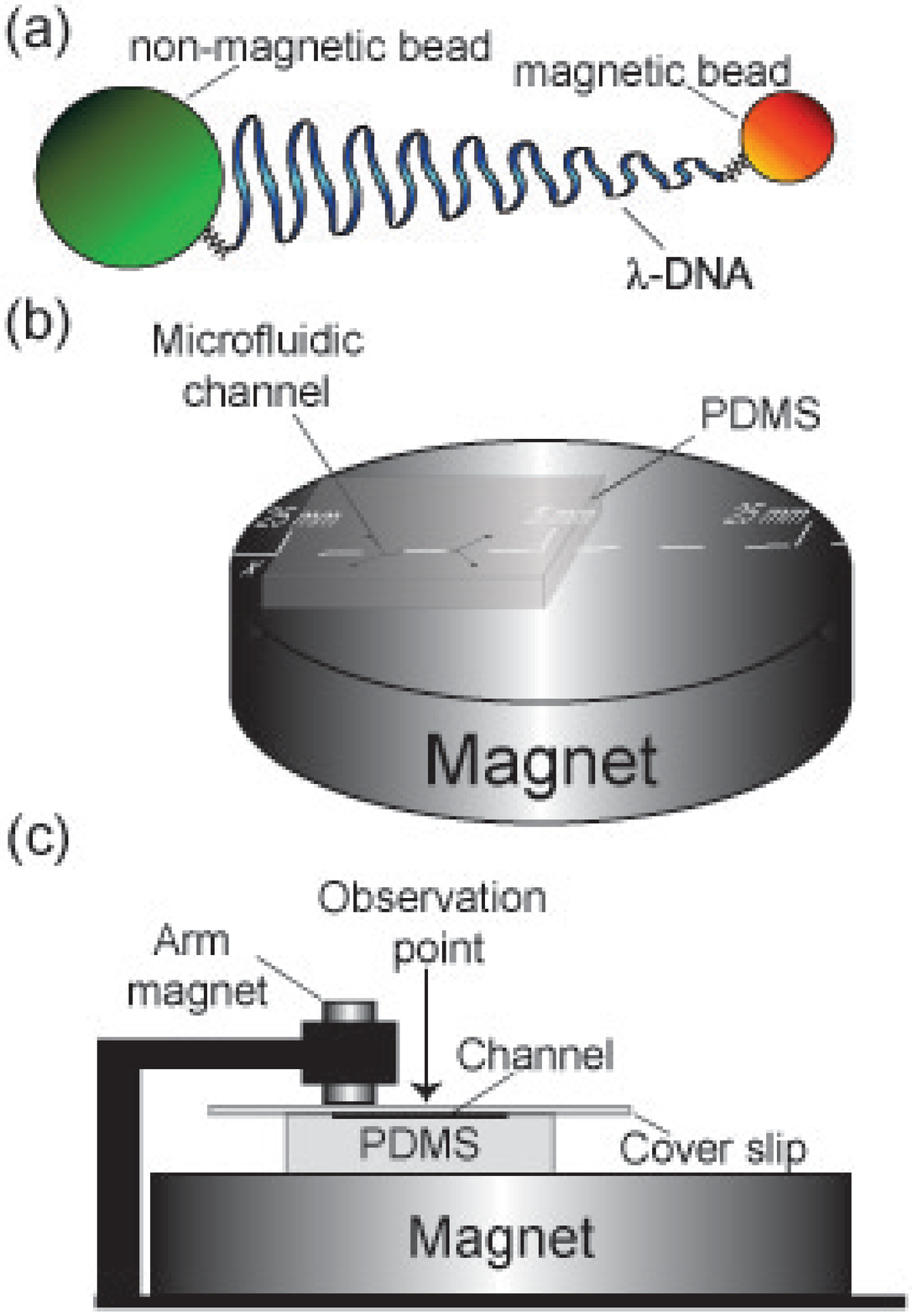}% size for 2column
\caption{\label{Image1} \textbf{(a)} Schematic illustration of a
dumbbell formed from a MB, a DNA and a NMB. \textbf{(b)} A disk
magnet with a microfluidic channel. The magnetic field is the
weakest in the center of magnet \mbox{$\mid\mathbf{B}\mid \approx
0.14$\,T} (position \mbox{$x=5$\,mm}) and the strongest
\mbox{$\mid\mathbf{B}\mid \approx 0.19$\,T}, at position
\mbox{$x=25$\,mm}. \textbf{(c)} The arm magnet is illustrated
together with the disk magnet and a microfluidic channel. The
observation point is roughly indicated.}
\end{figure}

An illustration of a dumbbell is shown on Fig.~\ref{Image1}(a).
The magnetic beads M-280 with diameters of \mbox{$2.8$\,$\mu$m}
and MyOne with diameters of \mbox{$1$\,$\mu$m} (Invitrogen, Oslo,
Norway) were investigated. The iron content for M-280 and MyOne
was \mbox{$\approx12$\,$\%$} and \mbox{$\approx24$\,$\%$},
respectively~\cite{Dynal}. In this study, we concentrated on
analyzing the results obtained with  M-280 MBs. The non-magnetic
beads (NMBs) are diamagnetic, polymer spheres with a diameter of
\mbox{$5.6$\,$\mu$m} (Bangs Laboratories, Inc., USA).

The flexible linker consisted of bacteriophage $\lambda$-DNA (New
England Biolabs, USA) with a contour length of
\mbox{$16.5$\,$\mu$m}. A biotinylated oligonucleotide was ligated
to the first recessed $3'$ end of \mbox{$\lambda$} and the
purified biotinylated DNA was then bound to streptavidin coated
M-280 MBs via a strong biotin-streptavidin bond. In order to
attach a streptavidin-coated NMB to the other terminus of the DNA
polymer, we used biotin incorporation via fill-in of the second
recessed $3'$ end with Klenow polymerase and biotin-dCTP. Finally,
the complete DB was formed by resuspending the reaction product in
aqueous buffer and adding streptavidin-coated NMBs. A full
description  of the dumbbell synthesis process will be published
elsewhere. Prior to magnetophoresis, the solutions were diluted in
deionized water containing \mbox{$1$\,$\%$} by weight
(\mbox{$1$\,wt\,$\%$}) of sodium dodecyl sulfate surfactant. This
minimizes nonspecific adhesive binding between the beads and the
surfaces of the channel.

All experiments on controlled movement and manipulation of the
dumbbells were done in microfluidic channels with widths from
\mbox{$200$\,$\mu$m} down to \mbox{$50$\,$\mu$m}. A mould with the
channel pattern was structured with SU-8 photoresist on a silicon
wafer using standard photolithography. The microfluidic device was
cast in polydimethylsiloxane (PDMS) and  subsequently sealed on a
glass cover slip~\cite{Whitesides}.

Magnetophoresis was performed initially on individual MBs and
subsequently the procedure was applied to the dumbbells. The
microfluidic chip was loaded with the MB suspension (typically
\mbox{$1$\,$\mu$l}) and subsequently transferred onto a
stationary, permanent disk magnet (rare earth magnet NdFeB,
\mbox{$50$\,mm} diameter, \mbox{$6$\,mm} width). An illustration
of the magnet with a microfluidic channel is shown in
Fig.~\ref{Image1}(b). The radial magnetic field gradient generated
by the disk shape (\mbox{$\nabla
\cdot\mathbf{B}\approx6.5$\,T/m}), enables us to achieve
directional movement of the beads along the channel in a
controlled manner (the $x$-direction was arbitrarily choosen). We
safely neglected effects of the gravitational force and thermal
energy~\cite{Fonnum}. The movement of individual MBs was observed
with a light microscope ($51$X, Olympus, Japan) equipped with a
$20\times$ objective (UPlan FI, Olympus, Japan). The trajectories
were recorded at different spatial positions on the magnet with a
digital camera (DP-70, Olympus, Japan). The integration time
varied between one second (strong magnetic field) up to ten
seconds (weak magnetic field). From these images one can calculate
the velocity of MBs, by dividing the recorded distance with the
integration time.

To estimate the dynamics of a DB in a microfluidic channel we
considered that the net force acting on a DB is composed of the
total force given by Eq.~\ref{totalforce}, acting on the MB
terminus, while the NMB experiences only the drag force.
Stretching of the DNA linker occurs only if sufficient force
difference is exerted on the DB between the MB and the NMB. The
dumbbell will move with equilibrium velocity approximating a value
that can be found from Eq.~\ref{totalforce} for the MB, and which
will determine the drag force acting on the NMB terminus. The
hydrodynamic drag force would act as a dynamic anchor, and at
equilibrium would be equal to the stretching force. For the disk
magnet, the stretching force was estimated
\mbox{$\approx0.1$\,pN}~\cite{forceestimate}, which was not
sufficient to completely stretch a DB. To increase the force, we
redesigned our magnetophoretic set up, introducing an additional
permanent magnet (\mbox{$4.5$\,mm} diameter, \mbox{$2.5$\,mm}
width), which can be moved at desired position over the stationary
disk magnet. A side view of a channel sandwiched between the arm
and disk magnet is shown in Fig.~\ref{Image1}(c). This provided a
significant increase in the magnetic field gradient experienced by
a MB. For example, even at a distance of \mbox{$2.5$\,mm} from its
center (the observation point in Fig.~\ref{Image1}(c)), the
stretching force is \mbox{$\approx0.5$\,pN}, which should be
sufficient to fully stretch a single strand of
$\lambda$-DNA~\cite{Bustamante, Chiou}.

%%%%%%%%%%%%%%%%%%%%%%%%%%%%%%%%%%%%%%%%%%%
% Results%
%%%%%%%%%%%%%%%%%%%%%%%%%%%%%%%%%%%%%%%%%%%

\begin{figure}
\includegraphics[width=75mm]{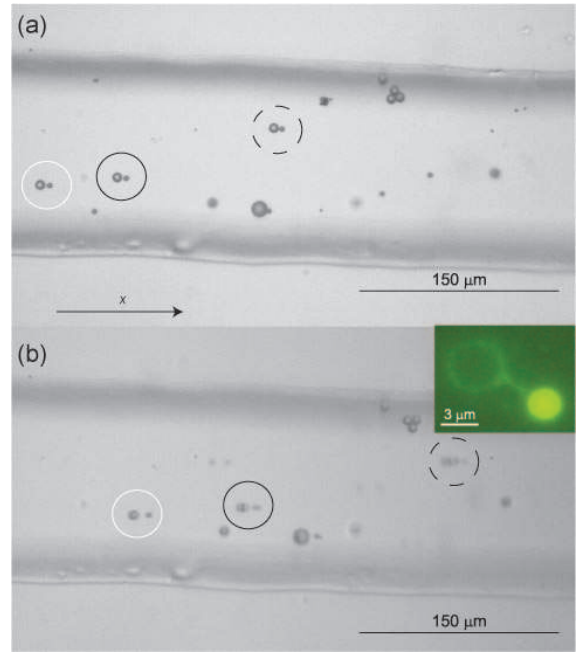}% size for 2column
\caption{\label{Image2} \textbf{(a)} An optical microscopy image
(\mbox{$20$\,$\times$} objective) of a microfluidic channel loaded
with DBs at relatively low strength magnetic field. The three
distinct DBs are indicated with circles (black, white and dashed
lines). The MBs are positioned on the right side of DB due to
increase in the gradient of magnetic field in the same direction.
\textbf{(b)} The same position as in (a) after locally increasing
the magnetic field and the gradient by the arm magnet. Stretched
DBs are indicated with the same circles as in (a). Inset: A
fluorescence image (oil immersion \mbox{$100$\,$\times$}
objective, integration time \mbox{$100$\,msec}), an individual,
partially stretched DB formed with a DNA as a link between the
beads. }
\end{figure}

Figure~\ref{Image2}(a) shows an optical image of a microfluidic
channel at position \mbox{$x=10$\,mm} on the disk magnet and
relatively far away from the arm magnet (\mbox{$>5$\,mm}). Three
partially stretched DBs are visible in the channel (indicated by
circles). The average stretching was found to be
\mbox{$1.3\pm0.3$\,$\mu$m}. Subsequently, we approached with the
arm magnet at a distance of \mbox{$\approx2.5$\,mm} from the
observation point. The image shown in Fig.~\ref{Image2}(b) is
taken at the same position of the microfluidic channel as in
Fig.~\ref{Image2}(a). The movements of all three DBs are clearly
visible despite the short integration time
\mbox{$\approx20$\,msec}. Moreover, the DBs from left to right
along the channel (the arm magnet is situated on the right side),
show an increase in velocity which is reflected in the image as
blurring (the faster the DB moves, the more pronounced the
blurring is). The average stretching is found to be
\mbox{$3.1\pm0.4$\,$\mu$m}, more then doubled compared to the
stretching obtained from Fig.~\ref{Image2}(a). This conformal
elongation of DB is significantly shorter then the full length of
$\lambda$-DNA. The main reason for such incomplete streching is
the presence of multiple DNA molecules in the linker since the
large surface area of the beads contained multiple streptavidin
molecules. In addition, some of the DNA molecules are very likely
to be wrapped around the beads, which further reduces the
probability of a full contour stretching of a DB. It is possible
to reduce multiple attachments of DNA by sonicating the samples
for few seconds (\mbox{$\approx4$\,sec}) which can lead to
shearing of DNA molecules. After sonication, the percentage of DBs
connected by a single strand was found to be around
\mbox{$20$\,$\%$}. However, the overall number of DBs was also
reduced, and at longer sonication times no DBs were visible, due
to complete shearing of the dumbbell linker. The fluorescent image
of a DB containing the beads and DNA is shown in inset of
Fig.~\ref{Image2}(b). The image is obtained with a florescence
microscope where the DNA is stained with a cyanine dimer dye,
YOYO-1 (Molecular Probes Inc., USA). The numerous DNA molecules
anchored at the edges of the beads are clearly visible as a bright
halo. The multiple DNA binding could be more controllably reduced
in the future by using magnetic nanoparticles which offer smaller
surface area per particle and hence fewer attachment sites.

\begin{figure}
\includegraphics[width=75mm]{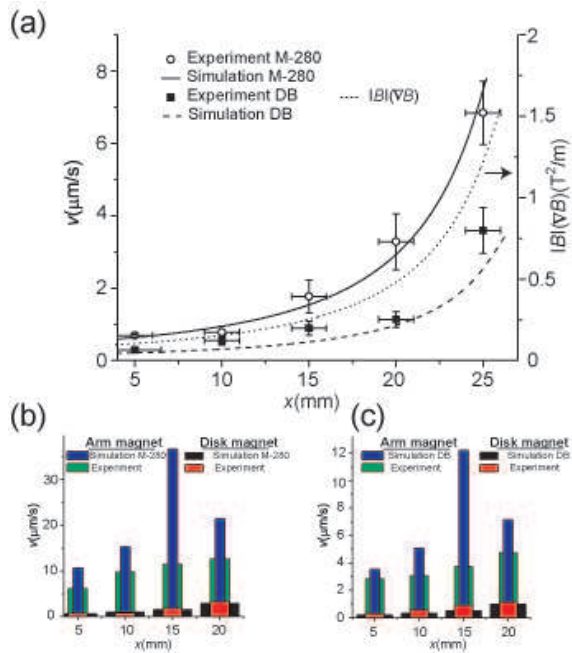}% size for 2column
\caption{\label{Image3} \textbf{(a)} Plot of velocities for the
MBs and DBs versus their spatial position on the disk magnet.
Symbols represent measured values, while full lines are simulated
velocities calculated by solving Eq.~\ref{totalforce} numerically.
The dashed line shows a dependance of product
\mbox{$\mid\textbf{B}\mid(\nabla \cdot \textbf{B}$}) along the
disk magnet (right axes). \textbf{(b)} and \textbf{(c)} Histograms
of measured and simulated velocities versus a spatial magnetic
position for the arm magnet and disk magnet. (b) corresponds to
MBs M-280 and (c) to DBs. }
\end{figure}

In order to model the system, we needed to know the magnetic field
and its gradient. The magnetic field was measured with a
gaussmeter (Lakeshore, 410). From the obtained values we were
numerically  able to extract the magnetic field gradient. In
Fig.~\ref{Image3}(a), we plotted the experimentally obtained
velocities at various spatial positions on the disk magnet for an
individual MB and a DB. The velocity-position dependance is
simulated by numerically solving Eq.~\ref{totalforce} using
\mbox{$2.8$\,$\mu$m} as the MB diameter (\mbox{$2.8$\,$\mu$m} and
\mbox{$5.6$\,$\mu$m} for a DB), the viscosity of water
\mbox{$\eta=8.9\cdot10^{-5}$\,kg\,m$^{-1}$s$^{-1}$}, the magnetic
permeability of vacuum \mbox{$\mu_0=4\pi\cdot10^{-7}$\,N/A$^2$},
while the magnetic susceptibility of the fluid comparative to the
MB was neglected. We would like to emphasize that the volume
magnetic susceptibility of the MB can be extracted as a fitting
parameter from the velocity-position measurements. For an
individual MB, we obtain \mbox{$\chi_b=0.13\pm0.04$} which is in
good agreement with previously reported values~\cite{Lee,
Amblard}. In the case of the DB, the dynamics of DNA were
neglected, which was justified due to the size of the used beads:
the drag force associated with the beads is greater than for a
single DNA molecule, multiple DNA molecules in the linker between,
or multiple DNAs wrapped around beads. From the plot in
Fig.~\ref{Image3}(a), we see that the experimental results for the
MBs and DBs are well described by this simple theoretical model.
As a qualitative check, we also display in Fig.~\ref{Image3}(a)
the product \mbox{$\mid \mathbf{B} \mid(\nabla \cdot\mathbf{B}$)}
versus spatial position on the disk magnet (dotted line). Clearly,
the \mbox{$\mid \mathbf{B} \mid(\nabla \cdot\mathbf{B}$)} product
is the main qualitative determinant for the dynamic behavior of
the objects in this study. Finally, with Fig.~\ref{Image3}(b) and
(c) we demonstrate the significant increase in the velocity and
stretching of a DB, when the arm magnet is used in combination
with the disk magnet. Fig.~\ref{Image3}(b) shows histograms of
experimentally found velocities, together with the corresponding
simulation for a single MB, while Fig.~\ref{Image3}(c) displays
similar histograms for the DB. Simulated velocities were extracted
by calculating the magnetic field and its gradient along the
length of the channel for the magnetic configuration illustrated
in Fig.~\ref{Image1}(c)~\cite{us1}. Both histograms in
Fig.~\ref{Image3}(b) and (c), display a marked increase in the
velocities of the test objects. While the overall trend of
velocity increase in the case of the arm magnet setup is present,
it is in lesser agreement with the simulation than in the case of
the bare disk magnet. We find that simulation values are highly
sensitive to the magnetic field gradient and the extracted values
strongly depend on the exact choice of observation position. Thus,
small inherent inaccuracies in precise determination of the
observation point position can lead to sizeable disagreement
between the model and the data.

%%%%%%%%%%%%%%%
% Conclusions %
%%%%%%%%%%%%%%%

In conclusion, we have investigated magnetophoresis of MBs and DBs
formed from magnetic and non-magnetic beads connected by DNA
strands. We find a strong dependence of dumbbell stretching on the
product of the magnetic field and its gradient. Experimentally
measured velocities of the MBs and DBs were compared with a
theoretical model, assuming simple equilibrium between magnetic
and drag force. We find solid agreement between the model and the
experimental data. This demonstrated approach to use a magnetic
force to stretch a flexible DNA-based dumbbell - with a
hydrodynamic drag force acting as a dynamic anchor on the
nonmagnetic terminus - has the potential to be a versatile tool in
future applied biological and biomedical separation devices.

%We demonstrated that by introducing additional arm magnet, further
%dynamical control and improvement are achieved.

\begin{acknowledgments}
We acknowledge experimental help from M. Hines and J.
Cooper-White, and valuable discussion with T. Meehan, R. Vogel and
M. Trau.
\end{acknowledgments}

% End of Main text
%---------------------------------------------------------------------------

%-------------------------------------------------------------

%-------------------------------------------------------------

\begin{thebibliography}{99}
\bibitem{forceestimate} Initial estimation of the magnetic force
was performed based on numerical simulation of the magnetic field
distribution calculated using Quick-field sofware.
\bibitem{us1} The magnetic field distribution is calculated with software package Quick field (not
shown), while all the other simulations are done using Mathematica
ver 5.1. The arm magnet data for the position \mbox{$x=25$\,mm} is
absent due to our magnetic design.
\end{thebibliography}
\end{document}